\documentclass[reprint,amsmath,amssymb,aps,prl]{revtex4-2}
\usepackage{xcolor}
\usepackage{tabularx}
\newcolumntype{Y}{>{\centering\arraybackslash}X}

\usepackage[newcommands]{ragged2e}
\usepackage{svg}
\usepackage{graphicx}
\usepackage{dcolumn}
\usepackage{bm}
\usepackage{chemformula} 
\usepackage[T1]{fontenc} 
\usepackage{comment}
\usepackage{physics}
\usepackage[utf8]{inputenc}
\usepackage[colorlinks=true,linkcolor=blue,filecolor=magenta,urlcolor=blue,citecolor=blue,pdfstartview=FitH]{hyperref}
\usepackage[justification=raggedright]{caption}
\usepackage{subcaption}
\usepackage{float}
\usepackage{blindtext}
\usepackage{array}
\newcolumntype{P}[1]{>{\centering\arraybackslash}p{#1}}
\usepackage{subfiles}
\usepackage{amsmath}
\usepackage{amsfonts}
\usepackage{amssymb}
\usepackage{multirow}
\usepackage{filecontents}
\definecolor{clr1}{rgb}{0,0,0}
\definecolor{clr2}{rgb}{0.0,0.35,0.0}
\newcommand{\figref}[1]{FIG.~\ref{#1}}
\newcommand{\eqaref}[1]{Eq.~\eqref{#1}}
\newcommand\apref{Appendix~\ref}
\newcommand\at[2]{\left.#1\right|_{#2}}

\newcommand{\tref}[1]{TABLE~\ref{#1}}
\newcommand{\beq}{\begin{equation}}
	\newcommand{\eeq}{\end{equation}}
\newcommand{\ie}{i.e.,}

\newcommand{\bse}{\begin{subequations}}
	\newcommand{\ese}{\end{subequations}}
\newcommand{\bea}{\begin{eqnarray}}
	\newcommand{\eea}{\end{eqnarray}}
\newcommand{\bem}{\begin{displaymath}}
	\newcommand{\eem}{\end{displaymath}}
\newcommand{\bmat}{\begin{bmatrix}}
	\newcommand{\ebmat}{\end{bmatrix}}

\newcommand{\bc}{\begin{center}}
	\newcommand{\ec}{\end{center}}

\begin{document}
	\title{Intermediate chiral edge states in quantum Hall Josephson junctions }
	
	\author{Partha Sarathi Banerjee}
	\affiliation{Department of Physics,
		Indian Institute of Technology Delhi,
		Hauz Khas, New Delhi 110016}

	\author{Rahul Marathe}
	\affiliation{Department of Physics,
		Indian Institute of Technology Delhi,
		Hauz Khas, New Delhi 110016}
	
	\author{Sankalpa Ghosh}
	\affiliation{Department of Physics,
		Indian Institute of Technology Delhi,
		Hauz Khas, New Delhi 110016}
	
	\date{\today}

\begin{abstract}
		A transfer-matrix-based theoretical framework is developed to study transport in superconductor-quantum Hall-Superconductor (SQHS) Josephson junctions modulated by local potential barriers in the quantum-Hall regime. The method allows one to evaluate the change in the conductivity of such SQHS Josephson junctions contributed by the intermediate chiral edge states (ICES) induced by these local potential barriers at their electrostatic boundaries at specific electron filling-fractions. It is particularly demonstrated how these ICES created at different Landau levels (LL) overlap with each other through intra- and inter-LL ICES mixing with the change in strength and width of the potential barriers. This results in different mechanisms for forming Landau bands when an array of such potential barriers are present. It is also demonstrated that our theoretical framework can be extended to study the lattice effect in a bounded domain in such SQHS Josephson junctions by simultaneously submitting the normal region to a transverse magnetic field and periodic potential.  
	\end{abstract}
	
	\maketitle
In superconductor-quantum Hall-superconductor (SQHS) Jospehson junctions (JJ) \cite{NORMAN199243,Ma_1993,Hatefipour2022, Vignaud2023, Akiho2024} where a quantum Hall (QH) system is proximity coupled to a superconductor (generally a $s$-type) on both sides, the Josephson effect occurs due to Andreev reflection in high magnetic fields \cite{Moore1999,Uhlishch2000,Hoppe2000,Zulicke2001,Eroms2005,Akhmerov2007, Takagaki2022} at the superconductor-Normal (SN) or superconductor-graphene (SG) interface. This was further theoretically explored in topological superconductors \cite{Chaudhary2020}, experimentally investigated in graphene \cite{Calado2015,Wang2021}, and other topological materials \cite{Tiwari2013}, in chiral Andreev edge states in twisted graphene bilayer and hBN encapsulated graphene \cite{Barrier2024, Zhao2020}, rhombohedral tetralayer graphen
\cite{Choi2025} to cite few examples. Such SQHS JJ are key building blocks for coherent superconducting quantum circuits \cite{Devoret1984,Likharev1985,Buttiker1987,Clarke1988} that can be used topological quantum information processing \cite{Stern2013, Beenakker2013, Clarke2014}. In a step that can significantly boost the understanding and manipulation of such SQHS JJ in a wider range of quantum transport,
in this work we show that local potentials in quantum Hall regime induce ICES (classically skipping orbits in the N region than at the Superconductor-quantum Hall (SQH) boundaries) at the electrostatic boundaries \cite{Halperin1982,Chklovskii1992} that significantly modify the band structure associated with the Landau levels (LL), and consequently alter the Josepshson conductivity through such junctions. 

In particular, we achieve two significant results: (a) 
We provide a method to calculate this variation of conductance in the ballistic regime due to induction of such ICES. In this method transfer matrices are derived from Blonder-Tinkham-Klapwijk (BTK)  \cite{Blonder1982} formalism which, in turn, gives dispersion, and subsequently conductivity is calculated using the Landauer-Buttiker formula. Further we demonstrate how this formalism can be extended to a periodically modulated SQHS JJ in the bounded N region subjected to a transverse magnetic field. 
(b) We analyze our results to establish two distinct mechanisms behind the formation of Landau bands from electron- and hole-LLs through intra- and inter-LL mixing of the ICES, and point out how they impact the Josephson conductivity in such junctions.

We theoretically model such SQHS JJ by considering the quantum Hall region of length $2L$ along $x$-direction, contains a finite number ($n_{B}$) of identical rectangular potential barriers with an uniform spacing between two consecutive barriers, and the vector potential corresponding to strong transverse magnetic field $B\hat{z}$ is taken in Landau gauge. A schematic of the system is given in \figref{disp1}(h). The imperfections in the SN and NS interfaces are modeled with two delta function potentials $U_1 \delta(x-L)$ and $U_2 \delta(x+L)$. The  S region, that can be created by suitably placed superconducting electrode on two-dimensional QH systems, occupies $\abs{x}>L$. All lengths are in the unit of magnetic length  $l=\sqrt{\hbar / \abs{eB}} = 25.6 /\sqrt{B}$ nm with  $B$ in Tesla, and energies are in the units of $\hbar \omega_{C}, \omega_{C}=\frac{eB}{m}$. For, $V(x)=\hbar \omega_C U(x)$(for details, see \apref{ap1}), the Bogoliubov-de-Gennes (BdG) equation that models different parts of the system in BTK formalism becomes: 

\begin{widetext}
	\begin{subequations}
	    \begin{align}
		\begin{bmatrix}
			-\frac{1}{2 } \frac{d^2}{dx^2} +F(x,+X) -\frac{\nu}{2} + U(x) && \Delta \\
			\Delta^* &&  \frac{1}{2 } \frac{d^2}{dx^2} - F(x,-X) + \frac{\nu}{2} - U(x)
		\end{bmatrix} \begin{bmatrix}
			f_X \\
			g_X
		\end{bmatrix}= E \begin{bmatrix}
			f_X \\
			g_X
		\end{bmatrix} \label{N}
	\end{align}
	
	\text{with,}\begin{equation} \label{pot}
		U(x) = U_0 \delta(x+L) + U_0 \delta(x-L) + \sum_{n=1}^{N} V_0 \Theta \left(x - x_n + \frac{d}{2}\right) \Theta \left(x - x_n - \frac{d}{2}\right)
	\end{equation}
	\end{subequations}
\end{widetext}

 Without the barrier \cite{Hoppe2000,JMF_1994}, the solutions of this BdG equation \eqaref{N},
with the electron and hole components of the wave-functions denoted by $f_X(x)$ and $g_X(x)$, are defined as:
\begin{figure*}
	\centering
	\includegraphics[width=\linewidth]{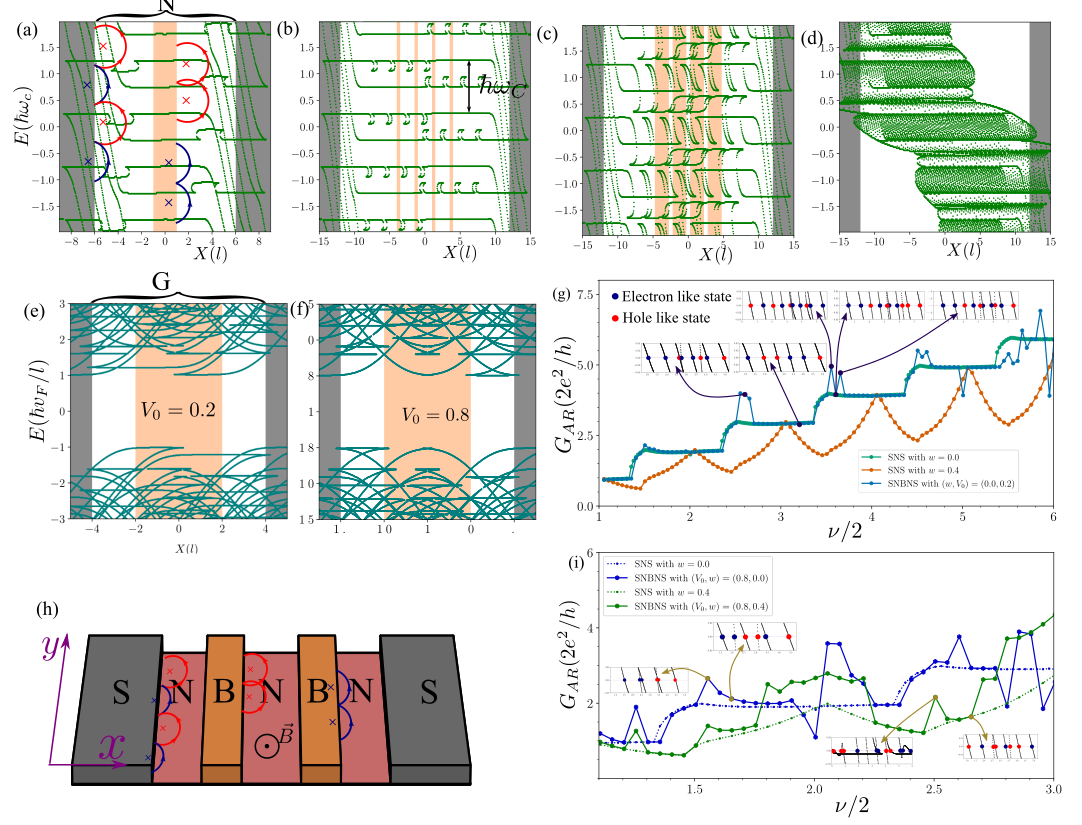}
	\caption{ \justifying We show the dispersion plots for the SNS junctions with barriers in the N region for $\nu = 5.5$ using \eqaref{nbarrier}. To show the presence of Landau levels, we set $\Delta_0 = 2.0 \hbar \omega_C$. In (a), we have one barrier with width $d=2$. In (b), (c) we have taken $4$ barriers in the N region with (b) $V_0 = 0.2$, separation $D = 2$ and $d= 1 $, (c) $V_0 = 0.9$, $D = 0.5$ and $d=3$ and in (d) we have taken $40$ barriers in the N region  with $d = 0.3$, $D = 0.3$. The red semicircles denote the classical electron orbit, and the blue semicircles denote the classical hole orbits.
    In (e) and (f), we show the dispersion for the monolayer graphene-based SQHS junction when the electrostatic barrier of height  $V_0= 0.2$ and $0.8$ is present in the QH region respectively. In (g), we compare the conductivity of the SNS junction with a single barrier in the N region with two cases of the SNS junction with $w=0$ and $w=0.4$. Here, $w$ is defined by $w=2 U_0/ \sqrt{\nu}$. In this case, we have taken $\Delta_0 = 0.01 \times \nu$. Here, the distance between two SN edges is taken as 6 and the width of the barrier is taken as 1. In the inset of (g) and (i) we show the intermediate chiral edge states which contribute to the fluctuation in conductivity. the red and blue dot denotes their electron like or hole like nature. The corresponding table of the hole probabilities are given in appendix. We have shown a schematic diagram of the model system that we are considering in (h). In (h) we have shown $2$ rectangular barriers in the N region.}
	\label{disp1}
\end{figure*}

\begin{subequations} \label{SNwf}
	\begin{align}
		f_X = a \chi_{\varepsilon_+} & = \begin{cases}
			a U \left( - \left\{ \frac{\nu}{2} + E\right\}, \sqrt{2} (x-X)\right) \text{	, in  N }  \\
			 d_- \gamma_{-} e^{i x k_-} + d_+ \gamma_{+} e^{-i x k_+} \text{	, in S} 
			\end{cases} \label{fX} \\
		g_X = b \chi_{\varepsilon_-} & = \begin{cases}
			b U \left( - \left\{ \frac{\nu}{2} - E\right\}, \sqrt{2} (x+X)\right) \text{	, in N}   \\
			 d_-  e^{i x k_-} + d_+  e^{-i x k_+} \text{	, in S} 
		\end{cases} \label{gX}
	\end{align}
\end{subequations}

Here, in the N region the solutions $U\left(- \frac{\nu}{2} \pm E, \sqrt{2}(x\mp X\right)$ are parabolic cylinder functions \cite{abramowitz1948} and $a$, $b$ and $d_{\mp}$ are constants. $X = p_y l^2/ \hbar$ is the guiding center co-ordinate of electrons and $p_y$ is the $y-$component of momentum. Also,

\begin{subequations}
	\begin{equation} \label{FX}
		F(x,X)=\begin{cases}
			\frac{X^2}{ 2} \text{, in the S region}  \\
			\frac{1}{2} (x - X)^2 \text{, in the N region} 
		\end{cases}
	\end{equation} 
	\begin{align}
		k_{\pm} & =  \left[ \left(\nu \pm i 2 \Delta_0\right) - X^2\right]^{1/2} \\
		\gamma_{\pm} & =   \frac{1}{\left(\frac{E_{nX}}{\Delta_0}\right) \mp \sqrt{\left(\frac{E_{nX}}{\Delta_0}\right)^2 -1}}
	\end{align}
\end{subequations}

In presence of the barriers, the problem can be approached in two different ways,
\textit{method-I:} we consider a finite number of electrostatic barriers in the N region with equal spacing between consecutive barriers, resulting in a potential profile given in \eqaref{pot} $V(x) = \sum_{n=1}^{n_{B}} V_0 \Theta \left(x - x_n + \frac{d}{2}\right) \Theta \left(x - x_n - \frac{d}{2}\right)$.Here, the $n^{th}$ barrier of width $d$ is centered at $x_n$, and the seperation $x_{n+1}-x_n=D$ .

The dispersion relation is derived from the transfer matrices that match the boundary conditions in every edge ((super-conductor-normal)SN, (normal-barrier)NB, (barrier-normal)BN and NS (normal-superconductor)) (for details see \apref{ap2}), and can be written as 

\begin{equation} \label{MM}
	\mathcal{M} = \begin{bmatrix}
		M_{SN} & 0 & 0 & \dots &0 &0 &0\\
		0 & M_{NB}^{(1)} & 0 &\dots &0 &0 &0\\
		0 & 0 & M_{BN}^{(1)} & \dots &0 &0 &0\\
		\vdots & \vdots &  \vdots & \ddots & \vdots & \vdots & \vdots\\
		0 & 0 & 0 & \dots & M_{NB}^{(n_{B})} & 0 & 0\\
		0 & 0 & 0 & \dots & 0 & M_{BN}^{(n_{B})}  & 0\\
		0 & 0 & 0 & \dots & 0 & 0  & M_{NS}
	\end{bmatrix}
\end{equation}

\begin{widetext}
	\beq \label{nbarrier}
	\det(\mathcal{M}) = \det(M_{SN}) \left[\prod_{i=1,2 \dots n_{B}} \det(M_{NB}^{(i)})  \det(M_{BN}^{(i)}) \right] \det(M_{NS}) =0 
 \eeq 
\end{widetext}
gives the dispersion. 
Each individual matrix in the diagonal of \eqaref{MM} is $(4\times 4 )$ matrix obtained from the boundary value conditions. 
A single potential barrier that is responsible for inserting ICES in the $N$ region can be written as, $V(x)=V_0 \Theta(x+d/2) \Theta(d/2-x)$ with 
\begin{equation} \label{bigM}
	\mathcal{M}= \begin{bmatrix}
		M_{SN} && 0 && 0 && 0 \\
		0 && M_{NB} && 0 && 0 \\
		0 && 0 && M_{BN} && 0 \\
		0 && 0 && 0 &&  M_{NS}
	\end{bmatrix}.
\end{equation}
Let us define two matrices, 
\begin{widetext}
		
	\begin{equation} \label{MX}
		M(X) = \left[ \begin{smallmatrix}
			\chi_{\varepsilon_+}(-X) && 0 && - \gamma_- && - \gamma_{+} \\
			0 && \chi_{\varepsilon_-}(-X) && 1 && 1 \\
			\at{\derivative{\chi_{\varepsilon_+}}{x}}{x=0} && 0 &&  -\gamma_{-}(ik_{-} + 2 U_0)  && \gamma_{+}(i k_{+} - 2 U_{0}) \\
			0 && \at{\derivative{\chi_{\varepsilon_-}}{x}}{x=0} &&  -(ik_{-} + 2 U_0) && \gamma_{+}(i k_{+} - 2 U_{0})
		\end{smallmatrix} \right] , M_B(X) = \left[ \begin{smallmatrix}
		\chi_{\varepsilon_+}(-X) && 0 && - \chi_{\varepsilon_{B+}}(-X) && 0 \\
		0 && \chi_{\varepsilon_-}(-X) && 0 && \chi_{\varepsilon_{B-}}(-X) \\
		\at{\derivative{\chi_{\varepsilon_+}}{x}}{x=0} && 0 && \at{\derivative{\chi_{\varepsilon_{B+}}}{x}}{x=0} && 0 \\
		0 && \at{\derivative{\chi_{\varepsilon_-}}{x}}{x=0} && 0 && \at{\derivative{\chi_{\varepsilon_{B-}}}{x}}{x=0}
	\end{smallmatrix} \right].
	\end{equation}
\end{widetext}
Now the $M_{SN}$ and $M_{NS}$ are derived from $M$ by axis transformations. For $M_{SN}$ the transformation is $x\rightarrow x+L$. For $M_{NS}$ the transformation is $x\rightarrow - x - L$. Now the $M_{BN}$ and $M_{NB}$ are derived from $M_B$ by axis transformations. For $M_{NB}$ the transformation is $x\rightarrow x+L_B$. For $M_{BN}$ the transformation is $x\rightarrow - x - L_B$. 

The above method can be easily generalized to potential's profile smoother than the rectangular barrier \cite{Bartos1994,Grover_2012} and unequal spacing.
\textit{Method-II:} In the second approach, initially the N region in the presence of transverse magnetic field is considered to be subjected to a one dimensional periodic potential of the form $V(x) = \sum_{n=- \infty}^{\infty} V_0 \Theta \left(x - n D + \frac{d}{2}\right) \Theta \left(x - n D - \frac{d}{2}\right)$ so that Bloch condition can be imposed. In the tight-binding limit this can be mapped to Hofstatder problem, but for an one-dimensional periodic potential.
The wavefunctions in $n$'th N region, extended from $x= x_n$ to $x_n + D$ is given by, 
\begin{equation} \label{nN1}
	(f,g)_{nX}  = (a,b)_n U\left[\left(-\frac{\nu}{2} \pm E\right), \sqrt{2} \left(x-x_n \mp X \right)\right].
\end{equation} 
Bloch condition gives 
\begin{equation} \label{bloch1}
	\frac{(a,b)_{n+1}}{(a,b)_n} = \cos( K_{(1,2)} (d+D))
\end{equation}
leading to Kronig-Penny type dispersion,
\begin{widetext}
		\begin{equation}
			\cos( K_{(1,2)} (d+D))  = \frac{U\left[\left(-\frac{\nu}{2} \pm E - V_0 \right), \sqrt{2} \left(d \mp X \right)\right]}{U\left[\left(-\frac{\nu}{2} \pm E\right), \sqrt{2} \left(\mp X \right)\right] } \times \frac{ U\left[\left(-\frac{\nu}{2} \pm E\right), \sqrt{2} \left(D \mp X \right)\right]}{  U\left[\left(-\frac{\nu}{2} \pm E - V_0 \right), \sqrt{2} \left( \mp X \right)\right]} \label{ebm}
		\end{equation}
\end{widetext}
where we calculate $K_{(1,2)}$ from \eqaref{eb} elaborated in \apref{ap3}. Under these conditions the allowed solutions are those $(E,X)$ values, for which one gets $\abs{\cos( K_{(1,2)} (d+D))}<1$ in \eqaref{ebm}.

\begin{figure*} 
    \centering
    \includegraphics[width=\linewidth]{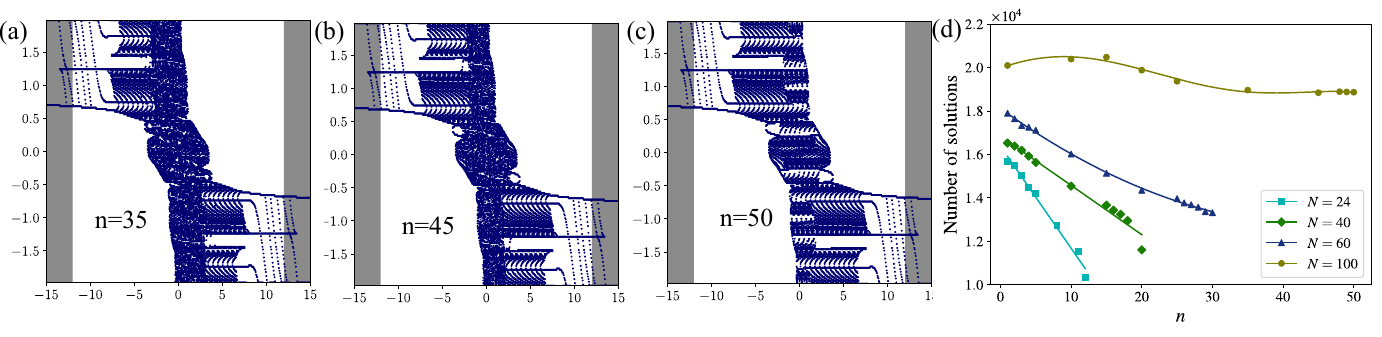}
    \caption{In (a), (b) and (c) we show the dispersion calculated from Method II, from \eqaref{nbarrierbloch} in the main text for $n_{B}=100$ barriers in the QH region of a SQHS junction. The $x$ and $y$ axes are same as in \figref{disp1} (a)-(d). In (a), (b) and (c) we have taken $n=35$, $n=45$ and $n=50$. $n=50$ gives same result as Method-I, shown in \eqaref{nbarrier} in the main text. In (c) we show the number of points in the $E$-$X$ dispersion plots (bound states) obtained from the method II \eqaref{nbarrierbloch} for the case of $n_{B}=24$, $40$, $60$ and $100$. The solid line denotes the fitted curve. The last points in each curve is where we do not have Bloch condition at all. This is same as the results obtained using Method-I.}
    \label{BlochFigure}
\end{figure*}

However, in the JJ,  $NS$ and $SN$ interfaces break the lattice translational symmetry. To introduce this aspect, we assume that the Bloch condition can be used with the following modifications. The change in the wavefunction is limited upto first $n$ barriers one the right and the left side (in reality this will be a limiting process) of the N region due to the presence of SN and NS edges, and inside that region the Bloch periodicity remains valid. Under this assumption, we increase $n$ in step of one, till we get covergence of the energy spectrum for $n$ and $(n+1)^{\text{th}}$ case.  Validity of this condition gets better and better as one approaches the limit  $\frac{n}{n_{B}} \ll 1$, which in turn implies  the stronger limit namely $(d+D)/L \ll 1$. This method has some similarity with laminar boundary layer theory \cite{Jog_2015,Anderson}
in fluid mechanics, however the context and the details of the methodology is completely different here. 
To support our argument that this method indeed allows one to implement the Bloch condition in a bounded region with necessary modifications, we compare the number of points in the $E$ vs $X$ dispersion plots (bound states) for different values of n using both methods. To that end, in \figref{BlochFigure} we show that as we change $n$, for large lattices the number of bound states (number of roots that gives the dispersion plot) starts showing convergence towards the result obtained from Method-I. Although, we do not discuss it here, a suitable measure to quantify this convergence can be established. As expected this shows that both methods lead to the same result for very large number of barriers. 
Unlike the method adopted by Hatsugai \cite{Hatsugai1993} the current scheme is not contingent upon the formation of magnetic Brillouin zone which is a consequence of  lattice translational symmetry. 

In this method, the dispersion can be obtained from,

\begin{widetext}
	\begin{align} \label{nbarrierbloch}
		\det(\mathcal{M}) = \det(M_{SN}) \left[\prod_{i=1,2 \dots n} \det(M_{NB}^{(i)})  \det(M_{BN}^{(i)}) \right] & \left[\prod_{j=1,2 \dots (n_{B}-2n)} \det(M_{NB,lat}^{(n+j)})  \det(M_{BN,lat}^{(n+j)}) \right] \notag \\
		& \left[\prod_{k=1,2 \dots n} \det(M_{NB}^{(N-n+k)})  \det(M_{BN}^{(N-n+k)}) \right] \det(M_{NS}) =0
	\end{align}
\end{widetext}

Here, first $n$ matrices $M_{NB}$ and $M_{BN}$ from both the left and right sides of the N region stay the same as in the case of method I. The Bloch condition is assumed for the rest of the matrices. Particularly (details in \apref{ap3}),

\begin{widetext}
\begin{equation} \label{barrierM}
	M_{B,lat}^{(n)}(X) = \left[ \begin{smallmatrix}
		e^{iK_1 (n-1)(D+d) }\chi_{\varepsilon_+}(-X) \!&\! 0 \!&\!  e^{iK_1 ((n-1)(D+d)+d) } \chi_{\varepsilon_{B+}}(-X) \!&\! 0 \\
		0 \!&\! e^{iK_2 (n-1)(D+d) } \chi_{\varepsilon_-}(-X) \!&\! 0 \!&\! e^{iK_2 ((n-1)(D+d)+d) }\chi_{\varepsilon_{B-}}(-X) \\
		e^{iK_1 (n-1)(D+d) } \at{\derivative{ e^{iK_1 x } \chi_{\varepsilon_+}}{x}}{x=0} \!&\! 0 \!&\! e^{iK_1 (n-1)(D+d) } \at{\derivative{\chi_{\varepsilon_{B+}}}{x}}{x=0} \!&\! 0 \\
		0 \!&\! e^{iK_2 (n-1)(D+d) }\at{\derivative{\chi_{\varepsilon_-}}{x}}{x=0} \!&\! 0 \!&\! e^{iK_2 ((n-1)(D+d)+d) }\at{\derivative{\chi_{\varepsilon_{B-}}}{x}}{x=0}
	\end{smallmatrix} \right].
\end{equation}

	\end{widetext}

The method of calculation of conductivity for such systems is described below. 

\begin{enumerate}
	\item Without any externally applied bias voltage, the Landau Level Andreev Bound States (LLABS) determined from the  $E=0$ with the dispersion obtained from \eqaref{nbarrier}
    , and the chemical potential is absorbed in filling fraction $\nu$ in the BdG equations of N and S regions (For details see \apref{ap1}) contributes to the Josephson current .  
	\item Assuming the transport is in the ballistic regime, in Landauer-Buttiker formalism \cite{Buttiker1988} the conductivity of this JJ can now be straight-forwardedly calculated by summing over the hole probability as  
	\begin{equation} \label{condbtk}
		G_{AR}=\frac{e^2}{\pi h} \sum_{j=1}^{n^*} B_j
	\end{equation}

    This summation runs on every point on the dispersion plot that intersects with $E=0$, as described in Step - 1. Each of these points correspond to a state with $(E,X)=(0,X_j)$. Here $B_j = \int_{x}\abs{g_{X_j}}^2$ of all electron-like states staisfying $\int_{x} \left[\abs{f_{X_j}}^2- \abs{g_{X_j}}^2\right]>0$ such that $B_j\le1/2$. Here $n^*$ is the total number of states for which we have $\int_{x} \left[\abs{f_{X_j}}^2- \abs{g_{X_j}}^2\right]>0$.
	\end{enumerate}

\begin{figure*}
  \centering
  \includegraphics[width=0.7\linewidth]{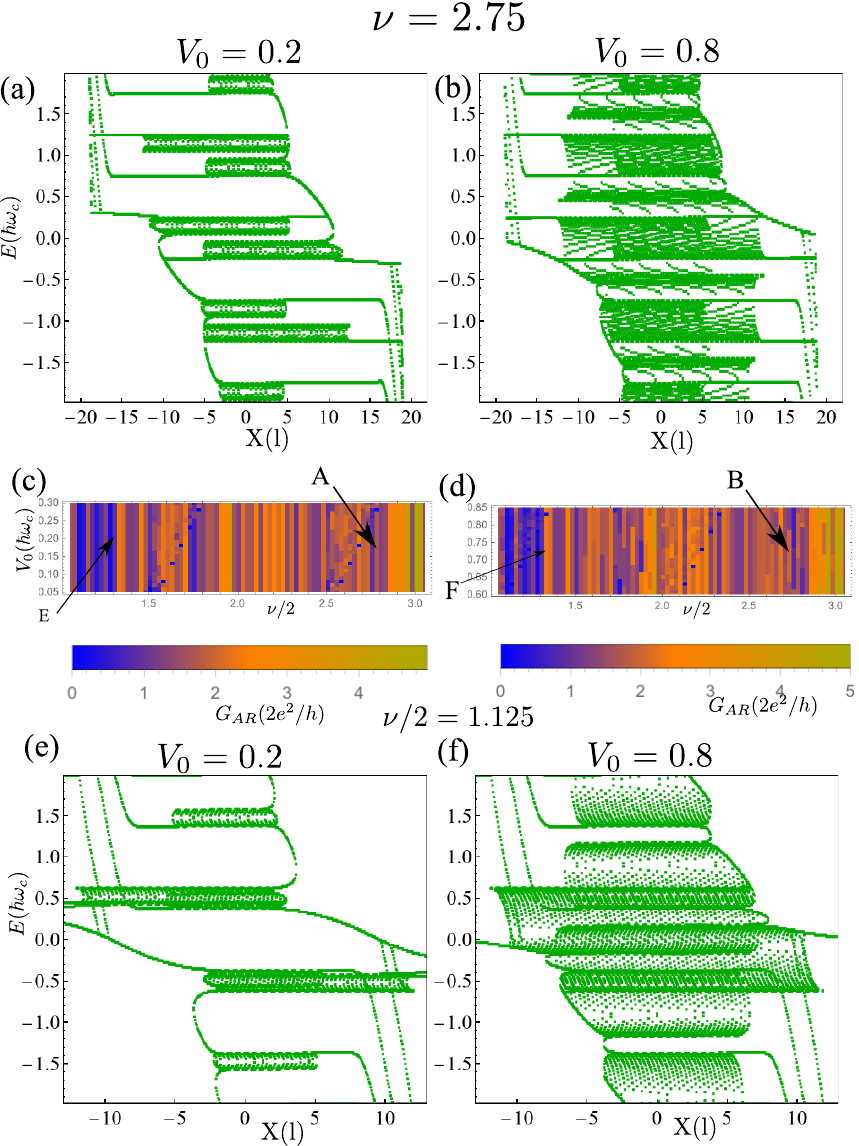}
  \caption{In (a), (b), (e) and (f) we show the dispersion plot for SN(BN)$^n_{B}$S junction with $n=30$. For (a) $\nu/2= 2.75$ and $V_0=0.2$, (b) $\nu/2= 2.75$ and $V_0=0.8$, (e)  $\nu/2= 1.125$ and $V_0=0.2$ and (f)  $\nu/2= 1.125$ and $V_0=0.8$. The with of the barriers are taken as $d=0.3$ and the separation between them is taken as $D=0.3$. In (c) and (d) we show the conductivity, calculated using \eqaref{condbtk} as a function of both $V_0$ and $\nu/2$. The $x$ axes of the dispersion plots (a), (b), (e) and (f) are $X$ and the $y$ axes are energy $E$. The $E$ vs $X$ dispersions (a), (b), (e) and (f) correspond to 4 points A, B , E and F in the conduction plots (c) and (d). For the dispersion we have taken $\Delta=2$ and the conduction plots we have taken $\Delta= 0.01 \nu$. The dispersion plots are calculated with same values of $\nu$ and $V_0$. However as we have changed the $\Delta$ for the conductance calculation the window of energy in which Andreev bound states are formed get reduced to $-0.01 \nu$ to  $+0.01 \nu$.}
  \label{densityplot}
\end{figure*}

In \figref{disp1} (a) -(d), using \eqref{nbarrier} we show the dispersion $E$ vs $X$ for (a) single barrier with width $d=2$, (b) four barriers with $d=1$, separation $D=2$ and $V_0=0.2$, (c)four barriers with $d=2$, separation $D=0.5$ and $V_0=0.9$ and (d) $40$ barriers with separation $D=0.3$ and $d=0.3$, with all barrier(s) placed symmetrically inside the QH region. The formation of the ICES and their variation in the electron and hole LLs in these figures are described in detail in \apref{degap} with the help of \figref{SNS} (a) and (b). Comparison of \figref{disp1}(a) and (b) shows that the addition of more electrostatic barriers in the QH region increases the number of intermediate chiral edge states inside the QH region. The intermediate chiral edge states form a convexo-concave structure over the usual LLs over a range of $X$. In (c) we reduced separation between the barriers $D$, and also increased the value of $V_0$. The increase in $V_0$ results in overlapping of convexo-concave (CC) structures from different LLs (inter LL overlap), whereas reduced  $D$ leads to overlapping of the adjacent CC structures in the same LLs (intra LL overlap). This profoundly impacts the subsequent LL band formation, when number of barriers are sufficiently large.  

In \figref{disp1}(d), a relatively large number of barriers $n_{B}=40$ in the QH region lead to the formation of LL bands due to inter and intra-LL overlap of the CC regions in the dispersion. The interband region also contain significant number of states. 
This plot allows us to understand the complexity involved in the formation of LL bands in such system. As one can see, the width of such LL bands, and the states in between such bands are different from that of an unbounded QH system in a one-dimensional lattice under tight-binding approximation \cite{Hatsugai1993}. In the subsequent discussion we shall explain this aspect in more detail.

The mixing between the Landau levels can also be understood by looking at the charge current. To complement the above discussion, we provide some representative plots in  \figref{current_dist} using \eqaref{chcr}
in \apref{AppI}.
The currents shown in \figref{current_dist} are currents of the lowest LLs to show the contribution of each LL. In an experimental system, the measured current is summed over all the LLs present in such system for a given JJ (with gap $\Delta_0$) with a phase difference ($\phi$). 
The plots demonstrate how with the reduction of the separation between barriers the current contribution due to LLs of a given quantum number interferes. It may be pointed out that the phase-dependent Josephshon current in such SQHS junction shows highly intriguing behavior that was predicted theoretically \cite{Ma_1993, Stone2011} and demonstrated in a recent experiment \cite{Vignaud2023}. In this work, we have not studied the effect of such local potential barriers on the phase-dependent Josephson current. 

For comparison we also provide the effect of a single barrier in QH region for superconductor-graphene-superconductor (SGS) junction in \figref{disp1}(e) and (f). The formation of ICES can also be seen here, though it is very different from the SNS junction due to the relativistic dispersion of charge career in graphene QH and consequent Dirac BdG quations. The theoretical analysis is provided in detail in \apref{apsgs}. Extension of this calculation for such SGS junction in QH regime in presence of multiple barrier will be treated in future work. 

Finally, in \figref{disp1} (g), (i) we plot the variation of Josephson conductivity as a function of filling fraction $\frac{\nu}{2}$ calculated with the help of \eqaref{condbtk}. 
In (g) particularly, we plot conductance for the cases of S-QH-S junctions (i) without any barrier, (ii) delta function barriers at the NS edges which affect the Andreev reflection and then (iii) scatterer in the form of one rectangular barrier symmetrically placed in the QH region which introduces ICES. In the inset we show the existence of the ICES at the Fermi energy, and its effect on the conductivity (see \tref{tab1}, \tref{tab2} and \tref{tab3} in \apref{tabhole}). 

In \figref{disp1}(i) we study the joint effect of a delta function scatterer at the edge of the NS (SN) regions, and a symmetrically placed rectangular barrier in the QH region, on the Jospephson conductance of the SQHS junction. Furthermore, the barrier height $V_0$ increases compared to \figref{disp1}(g). A pronounced change in conductance was observed. 

To gain more insight in the modification of Jospephson conductivity due to the existence of such ICES and the consequent LL band formation, in 
\figref{densityplot} (c) and (d) we study the variation of conductace due to multiple equidistant rectangular barriers again symmetrically placed in the QH region over a range of $\nu$ and $V_0$. To elucidate the resulting behavior further, we compare the dispersion relation in (a), (b), (e) and (f) which correspond's to four specific points in the $V_{0}, \nu$ plane. We can observe strong Josephson conductance fluctuations over the plataues, seen in \figref{disp1} (i), (g). The fluctuation is due the formation of Landau bands through the intra and inter LL overlap of ICES. 
In the first case, for lower value of $V_{0}$, the dispersion in \figref{densityplot} (a) and (e) shows that the band formation due to intra LL overlap of the ICES. 
There are large gaps between the bands and clear edge states between the bands. This is  somewhat similar to the case studied in the seminal work by Hatsugai \cite{Hatsugai1993} with two fundamental differences (a) broken lattice translational symmetry by NS interface  and (b) electron hole conversion due to Andreev reflection.
In principle, those results generalized for 
electron and hole LLs can be recovered by performing calculation 
within our framework in the limit $\lim_{(V_{0}\rightarrow \infty, d \rightarrow 0)}V_{0}d =c~\text{(finite~constant)}$ and number of barriers $n \rightarrow \infty$, and making necessary tight binding approximations. 

The second case refers to the dispersion plotted in \figref{densityplot} (b) and (f). Here the LL bands are more complex and accompanied by additional inter LL overlapping of ICES that mixes different quantum numbers corresponding to particle and hole. This mechansim of formation of LL bands is different from the previous one, and also contains a significant number of intermediate states between two successive LL bands.  We provide in \apref{degap} with \figref{SNS}, a detailed explanation of the differences between the two different mechanisms with the help of dispersion relation for one bariier case, but for two different values of $V_{0}$. Clear distiction between these two different mechanisms of formation of LL bands due to intra and inter LL overlap of ICES injected by array of rectangular barrier potentials and the modification of Josephson conductivity in S-QH-S junction is one of most important findings of this work.

Any such SQHS JJ will inherently contain local potential barriers. Our work thus, not only provide a clear theoretical understanding of the variation of the Jospehson conductivity 
through the inter and intra LL network of edge states induced by these local potentials, but also provide a clear roadmap towards manipulation of the device performance using this analysis. It clearly shows how these barriers can be used to modulate such conductivity, and provide insight in the formation of Landau bands through different mechansim involving the interference of such edge states. 
Given the experimental demonstration of band conductance oscillation in graphene superlattice \cite{Huber2022,Mreńca-Kolasińska2023}, re-entrant superconductivity in Quantum Hall regime \cite{Maska2002,RevModPhys.64.709,Chaudhary2021, Shaffer2021, Vishwanath2025}, spectroscopy of fractal Hofstatder spectrum \cite{Nuckolls2025} in recent work, our method can be used to understand the impact of ICES in junctions made with such lattice QH systems. The impact of such local potential barriers or their arrays on the possibility of observation of the elusive Majorana
fermions \cite{Nilsson2008,Qi2010,FuMaj,Beenakker2013} in the chiral edge states of such SQHS junction can also be studied in future using the theoretical framework developed in this paper. Extension of this work to junctions made of multilayer graphene, twisted hetero-structures, understanding the impact of local barriers on current-phase relation in such juctions will significantly impact the insertion of such SQHS JJ in superconducting quantumn circuits and consequent quantum information processing.  

We thank U. Zuelicke and D. Aggarwal for helpful discussion during very early stage of this work. PSB was supported by a MHRD fellowship. SG was partially supported by a MFIRP project (MI02545G) of IIT Delhi in the initial part of this work. 
\appendix
\setcounter{table}{0}
\renewcommand{\thetable}{\Alph{section}\arabic{table}}
\setcounter{figure}{0}
\renewcommand{\thefigure}{\Alph{section}\arabic{figure}}

\section{Derivation of BdG equation for the S, B and N regions} \label{ap1}
The BdG equation in SQHS junction with a potential can be written as (details in \cite{Parthathesis}),
\beq \label{ham}
	\begin{bmatrix}
		H_0(x,X) -\mu && \Delta \\
		\Delta^* && \mu - H_0(x,-X)
	\end{bmatrix} \begin{bmatrix}
		f_X \\
		g_X
	\end{bmatrix}= E \begin{bmatrix}
		f_X \\
		g_X
	\end{bmatrix} 
\eeq
For a uniform magnetic field $\mathbf{B} = B \hat{z}$, in Landau gauge the vector potential becomes, $\mathbb{A}= x B \hat{y}$ in the N and B region and $0$ in the S region. Unit of length is magnetic length ($l = \sqrt{\frac{\hbar}{\abs{eB}}}$) and energy in the units of $\hbar \omega_C$, chemical potential (filling fraction) $\nu = 2 \mu / \hbar \omega_C$. This leads to
\beq \label{hamfg}
    H_0(x,X) = \frac{\left(\mathbf{p} - e \mathbb{A} \right)^2}{2 m} + V(x)=-\frac{1}{2 } \frac{d^2}{dx^2} + F(x,X) + U(x)
\eeq
Here, $F(x,X)$ is defined in \eqaref{FX} in the main text.
The pair potential for an NS junction can be taken as,
\begin{equation}
	\Delta=\begin{cases}
		\Delta_0 \text{, in S region}\notag \\
		0 \text{, in B and N region}
	\end{cases}
\end{equation}
The S region is defined as $\abs{x}>L$.  
By putting the form of $H_0(x,X)$ given in \eqref{hamfg}
in \eqaref{ham}, we get \eqaref{N} in the main text.

Since in  the normal region,  $\Delta=0$, \eqaref{ham} decoupled into following equations,
\begin{align}
    \left[H_0(x,X) -\mu \right] f_X & = E f_X \label{a3}\\
    \left[H_0(x,-X) -\mu \right] g_X & = -E g_X \label{a4}
\end{align}

\begin{figure*}
    \centering
    \includegraphics[width=\linewidth]{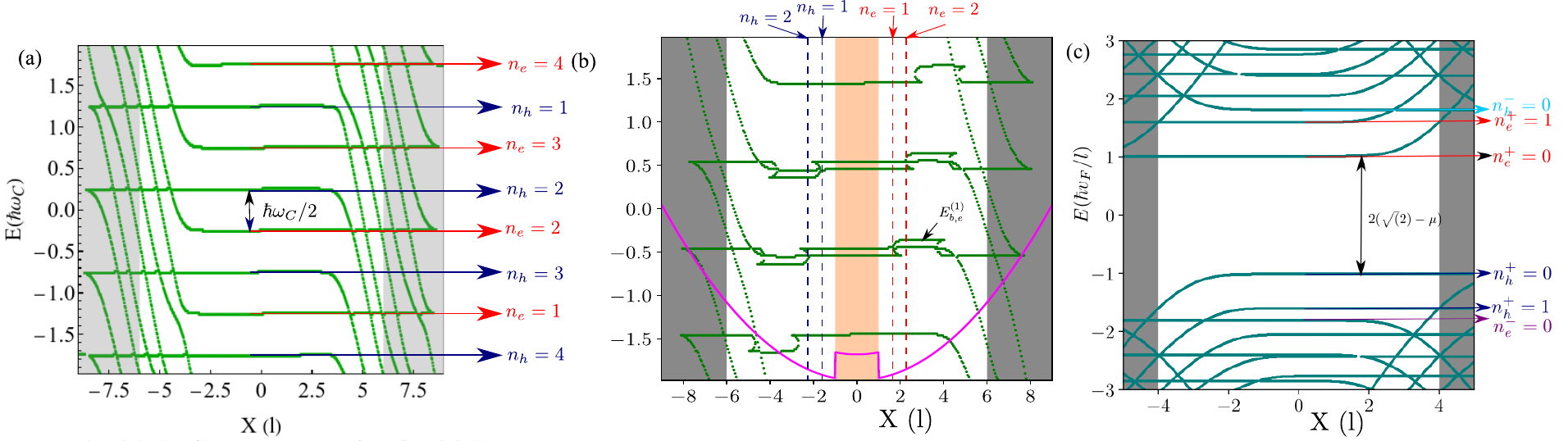}
    \caption{In (a) we show the dispersion(energy vs guiding center) plot for an SNS junction for the case of $\nu=5.5$. These are the solutions of $\det(M_{SN})\det(M_{NS})=0$. We denote the LLs from $f_X$ by $n_e$ and  LLs from $g_X$ by $n_h$. In (b) we show the dispersion for the case of a single barrier present in the QH region. We have taken $\nu=2.1$ and $d=2$. The maroon curve shows the effective potential $V_{eff}= (x-X)^2 /2 + V_0 \Theta (-x+d/2) \Theta(x+d/2)$ acting on the LLs for $X=0$. The value of the y axis in this curve is scaled appropriately to display with the dispersion plot and the $x$ axis is exact. The dotted red (blue) lines show the positions of the start of the lifting of degeneracy in $f_X$ ($g_X$) LLs due to barrier potential in the QH region. $n_e$ and $n_h$ are the same as in (a). $E_{b,e}^{(1)}$ corresponds to the eigenvalue inside the barrier region, and we have given its corresponding expression in the text. In (c), we show  the dispersion of the monolayer graphene-based SQHS junction with the separation between the two superconductors $2 L=8$, $\mu=0.4$ and superconducting gap $\Delta_0=10$. Here we have measured the lengths in the units of magnetic length $l$ and energies in the units of $\frac{\hbar v_F}{l}$.}
    \label{SNS}
\end{figure*}

Wavefunctions $f_X$ and $g_X$  from \eqaref{a3} and \eqref{a4} are the solutions of same equation \eqref{a3} with transformation $E\rightarrow -E$ and $X \rightarrow -X$.

\begin{align}
    H_0(x,X) f_X & = (E+\mu) f_X \label{a5}\\
    H_0(x,-X)  g_X & = -(E-\mu) g_X \label{a6}
\end{align}

In \figref{SNS}, we plot $E$ vs $X$ for these solutions. If we confine one dimensional electrons like \cite{Halperin1982}, we get the solutions of \eqaref{a5}. However, here the presence of $f_X$ and $g_X$ give two types of Landau levels with energies  $(E+\mu)$ and $-(E-\mu)$ and underscores the effect of Andreev refelction on the Landau levels in a SNS junction. From \eqaref{a5} and \eqref{a6}, the $n$'th LL from $f_X$ and $g_X$ has energy eigenvalues, $\left[ (n_e+ 1/2)  - \nu/2 \right]$ and $-\left[ (n_e+ 1/2)  - \nu/2 \right]$ respectively. In \figref{SNS} we see the energy gap is between the states $n_e=2$ and $n_h=2$. Their energy gap becomes $\hbar \omega_C/2$. 
\section{Boundary value conditions and the transfer matrices} \label{ap2}

For an NS junction, the wavefunctions, $f_X$ and $g_X$ are shown in \eqaref{SNwf}. The boundary value conditions are that the wavefunctions  should match at the boundary and the discontinuity of the derivative of the wavefunction is adjusted by the delta potential (shown in \eqaref{pot}) at the SN boundary. Using this, the boundary value conditions for the SN boundary at $x=0$ is given by

\begin{widetext}
    \begin{subequations}  \label{bvcq1}
	\begin{align}
		a \times \chi_{\varepsilon_+}(-X) + b\times0 &= d_-\times \gamma_{-}+ d_+\times\gamma_{+} \\
		a \times 0 + b\times \chi_{\varepsilon_-}(X) &= d_- + d_+ \\
		a \times \at{\derivative{\chi_{\varepsilon_+}}{x}}{x=0}- d_-\times \gamma_{-} \times (i k_-) -d_+\times \gamma_{+} \times (-i k_+)&=2 U_0 \left(d_-\times \gamma_{-}+ d_+\times\gamma_{+}\right) \\
		b \times \at{\derivative{\chi_{\varepsilon_-}}{x}}{x=0}- d_-\times  (i k_-) -d_+\times  (-i k_+)&=2 U_0 \left(d_- + d_+\right)
	\end{align}
\end{subequations}
\end{widetext}

The coefficients of $a$, $b$, $d_-$ and $d_+$ from this boundary value equation gives the elements of the transfer matrix $M(X)$ in \eqaref{MX}. In the SNS junction that we are considering, there the SN and NS boundaries are at $x=-L$ and $+L$ respectively. If we apply the the boundary values in the similar to Eqs. \eqref{bvcq1} and get the coefficients we get matrix $M_{SN}$ and $M_{NS}$ in  \eqaref{bigM}, as mentioned in the main text. 

We calculate the barrier boundary conditions at $x=0$, where the left side is N region and right side is B region. Then the boundary value conditions are given by,

\begin{subequations} \label{bvcbarrier}
    \begin{align} 
    a \times \chi_{\varepsilon_+}(-X) + b \times 0 &= a_B \times \chi_{B,\varepsilon_+}(-X) + b_B \times 0 \\
    a \times 0 + b \times \chi_{\varepsilon_-}(-X) &= a_B \times 0 + b_B \times \chi_{B,\varepsilon_-}(-X) \\
    a \times \at{\derivative{\chi_{\varepsilon_+}}{x}}{x=0} + b \times 0 &= a_B \times \at{\derivative{\chi_{B,\varepsilon_+}}{x}}{x=0} + b_B \times 0 \\
    a \times 0 + b \times \at{\derivative{\chi_{\varepsilon_-}}{x}}{x=0} &= a_B \times 0 + b_B \times \at{\derivative{\chi_{B,\varepsilon_-}}{x}}{x=0}
\end{align}
\end{subequations}

This coefficients of $a$, $b$, $a_B$ and $b_B$ in the boundary value conditions Eqs. \eqref{bvcbarrier} give us the matrix $M_B(X)$ in \eqaref{barrierM} in the main text. Now in similar way, if we calculate the boundary value conditions for the other NB and BN boundaries located at $x_n$ in \eqaref{pot}, we get all the $M_{NB}^{(n)}$ and $M_{BN}^{(n)}$ in \eqaref{MM}.

\setcounter{figure}{0}
\section{Derivation of $M_{B,lat}^{(n)}$ satisfying Bloch condition} \label{ap3}

The wavefunctions in $n^{\text{th}}$ N region, extended from $x= x_n$ to $x_n + D$ is given by, 

\begin{equation} \label{nN}
	(f,g)_{nX}  = (a,b)_n U\left[\left(-\frac{\nu}{2} \pm E\right), \sqrt{2} \left(x-x_n \mp X \right)\right]
\end{equation} 

The wavefunctions in $n^{\text{th}}$ B region, extended from $x= x_n+ D$ to $x_n + D+ d$ is given by, 

\begin{align}
    & (f,g)_{B, nX}   \notag\\
    & = (a,b)_{B,n} U\left[\left(-\frac{\nu}{2} \pm E - V_0 \right), \sqrt{2} \left(x-x_n - D \mp X \right)\right] \label{nB}
\end{align} 
	

The wavefunctions in $(n+1)$'th N region, extended from $x= x_n + D + d + D $ to $x_n + 2 D + 2 d$ is given by, 

\begin{align} 
	& (f,g)_{(n+1)X} = \notag \\
    & (a,b)_{n+1} U\left[\left(-\frac{\nu}{2} \pm E\right), \sqrt{2} \left(x-x_n - D -d \mp X \right)\right] \label{np1N}
\end{align}

Now the boundary value condition in $x= x_n+D$ gives,

\begin{widetext}
	\begin{align}
		a_n U\left[\left(-\frac{\nu}{2} + E\right), \sqrt{2} \left(D - X \right)\right] & = a_{B,n} U\left[\left(-\frac{\nu}{2} + E - V_0 \right), \sqrt{2} \left( - X \right)\right] \\
		b_n U\left[\left(-\frac{\nu}{2} - E\right), \sqrt{2} \left(D + X \right)\right] & =  b_{B,n} U\left[\left(-\frac{\nu}{2} - E - V_0\right), \sqrt{2} \left(+ X \right)\right] 
	\end{align}
\end{widetext}

This boundary condition gives,

\begin{equation}
	\begin{bmatrix}
		a_{B,n} \\ b_{B,n}
	\end{bmatrix} = 
    {
	\renewcommand{\arraystretch}{0.5} 
	\setlength{\arraycolsep}{0.5pt}
    \left[\begin{smallmatrix}
	\frac{ U\left[\left(-\frac{\nu}{2} + E\right), \sqrt{2} \left(D - X \right)\right]}{  U\left[\left(-\frac{\nu}{2} + E - V_0 \right), \sqrt{2} \left( - X \right)\right]} & 0 \\
	0 & \frac{U\left[\left(-\frac{\nu}{2} - E\right), \sqrt{2} \left(D + X \right)\right]}{  U\left[\left(-\frac{\nu}{2} - E - V_0\right), \sqrt{2} \left(+ X \right)\right] }
	\end{smallmatrix}  \right]
    }
    \begin{bmatrix}
	a_{n} \\ b_{n}
	\end{bmatrix}
\end{equation}

Again the boundary value condition at $x = x_n + D + d$ gives,

\begin{widetext}
	\begin{align}
		a_{B,n} U\left[\left(-\frac{\nu}{2} + E - V_0 \right), \sqrt{2} \left(d - X \right)\right] & = a_{n+1} U\left[\left(-\frac{\nu}{2} + E\right), \sqrt{2} \left(- X \right)\right] \\
		b_{B,n} U\left[\left(-\frac{\nu}{2} - E - V_0\right), \sqrt{2} \left(d + X \right)\right] & = b_{n+1} U\left[\left(-\frac{\nu}{2} - E\right), \sqrt{2} \left(+ X \right)\right]
	\end{align}

This boundary condition gives,

\begin{equation}
	\begin{bmatrix}
		a_{n+1} \\ b_{n+1}
	\end{bmatrix} = \left[ \begin{smallmatrix}
		\frac{U\left[\left(-\frac{\nu}{2} + E - V_0 \right), \sqrt{2} \left(d - X \right)\right]}{U\left[\left(-\frac{\nu}{2} + E\right), \sqrt{2} \left(- X \right)\right] } & 0 \\
		0 & \frac{U\left[\left(-\frac{\nu}{2} - E - V_0\right), \sqrt{2} \left(d + X \right)\right]}{U\left[\left(-\frac{\nu}{2} - E\right), \sqrt{2} \left(+ X \right)\right]}
	\end{smallmatrix} \right] \left[\begin{smallmatrix}
		a_{B,n} \\ b_{B,n}
	\end{smallmatrix}\right]
\end{equation}

\end{widetext}

\begin{figure*}
	\centering
	\includegraphics[width=\linewidth]{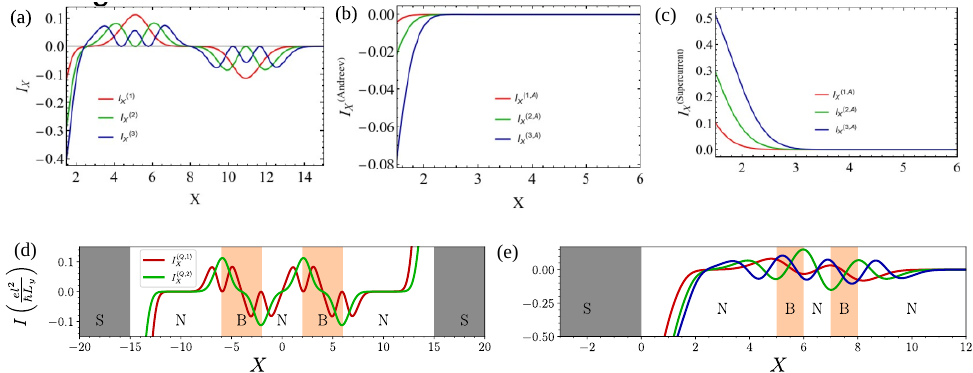}
	\caption{(a) Edge current, (b) Andreev current and (c) Supercurrent calculated from \eqaref{chcurr} for the first three electron-like LL states for an NS junction with a barrier in the N region. In (d), we show the charge current from \eqaref{chcr} for an SNS junction with two barriers of $d=4$ and $D=4$. In (e), we show the charge current for $d=1$ and $D=1$ in an NS junction with two barriers in the N region.}
	\label{current_dist}
\end{figure*}

By combining this we get,

\begin{widetext}
	\begin{equation} \label{blochtm}
		\begin{bmatrix}
			a_{n+1} \\ b_{n+1}
		\end{bmatrix} = \begin{bmatrix}
		\frac{U\left[\left(-\frac{\nu}{2} + E - V_0 \right), \sqrt{2} \left(d - X \right)\right]}{U\left[\left(-\frac{\nu}{2} + E\right), \sqrt{2} \left(- X \right)\right] } & 0 \\
		0 & \frac{U\left[\left(-\frac{\nu}{2} - E - V_0\right), \sqrt{2} \left(d + X \right)\right]}{U\left[\left(-\frac{\nu}{2} - E\right), \sqrt{2} \left(+ X \right)\right]}
		\end{bmatrix} \begin{bmatrix}
		\frac{ U\left[\left(-\frac{\nu}{2} + E\right), \sqrt{2} \left(D - X \right)\right]}{  U\left[\left(-\frac{\nu}{2} + E - V_0 \right), \sqrt{2} \left( - X \right)\right]} & 0 \\
		0 & \frac{U\left[\left(-\frac{\nu}{2} - E\right), \sqrt{2} \left(D + X \right)\right]}{  U\left[\left(-\frac{\nu}{2} - E - V_0\right), \sqrt{2} \left(+ X \right)\right] }
		\end{bmatrix} \begin{bmatrix}
		a_{n} \\ b_{n}
		\end{bmatrix}
	\end{equation}
\end{widetext}

As we are putting a periodic potential $V(x) = \sum_{n=- \infty}^{\infty} V_0 \Theta \left(x - n D + \frac{d}{2}\right) \Theta \left(x - n D - \frac{d}{2}\right)$, in \eqaref{hamfg}. For an infinitely large lattice, the wavefunctions, $f_X$ and $g_X$ in \eqaref{nN} and \eqaref{np1N} satisfies the Bloch conditions. Under the bloch condition if we define the Bloch momentums of electrons and holes as $K_1$ and $K_2$,

\begin{equation} \label{bcc}
    (f,g)_X(x+d+D) = (f,g)_X(x) \times cos( K_{1,2} (d+D))
\end{equation}

In \eqaref{blochtm}, we see that the ratio of the wavefunctions are real numbers. So we put the Bloch condition with only cosine term. Here, we have separated the electron and hole Bloch momentum as they are decoupled in N and B region. We satisfy the Bloch condition saparately for electron and hole part of the wavefunctions, which are stationary solutions of \eqaref{nN}. \eqaref{bcc} gives,

\begin{equation} \label{bloch}
	\frac{(a,b)_{n+1}}{(a,b)_n} = \cos( K_{(1,2)} (d+D))
\end{equation}

Combining \eqaref{blochtm} and \eqref{bloch}, we get

\begin{widetext}
\begin{subequations} \label{blochcondition}
    	\begin{align}
		\cos( K_{1} (d+D)) & = \frac{U\left[\left(-\frac{\nu}{2} + E - V_0 \right), \sqrt{2} \left(d - X \right)\right]}{U\left[\left(-\frac{\nu}{2} + E\right), \sqrt{2} \left(- X \right)\right] } \times \frac{ U\left[\left(-\frac{\nu}{2} + E\right), \sqrt{2} \left(D - X \right)\right]}{  U\left[\left(-\frac{\nu}{2} + E - V_0 \right), \sqrt{2} \left( - X \right)\right]} \label{eb}\\
		\cos( K_{2} (d+D)) & = \frac{U\left[\left(-\frac{\nu}{2} - E - V_0\right), \sqrt{2} \left(d + X \right)\right]}{U\left[\left(-\frac{\nu}{2} - E\right), \sqrt{2} \left(+ X \right)\right]} \times \frac{U\left[\left(-\frac{\nu}{2} - E\right), \sqrt{2} \left(D + X \right)\right]}{  U\left[\left(-\frac{\nu}{2} - E - V_0\right), \sqrt{2} \left(+ X \right)\right] } \label{hb}
	\end{align}
\end{subequations}
\end{widetext}

\eqaref{blochcondition} is same as \eqaref{ebm} in the main text. Now for a large number of barriers, the allowed solutions are those $(E,X)$ values, for which one gets $\abs{\cos( K_{1} (d+D))}<1$ in \eqaref{eb} and $\abs{\cos( K_{2} (d+D))}<1$ in \eqaref{hb}. Now we have the superconducting boundaries of this lattice at $x= -L$ and $x=+L$. We assume the solutions are Bloch periodic inside this region. Outside the region it take the usual superconductor wavefunction defined in \eqref{fX} and \eqref{gX}.

Now, with the wavefunctions $f_X$ and $g_X$ described in \eqaref{nN} and inserting the Bloch conditions \eqaref{blochcondition}, we use the boundary value conditions similar to \apref{ap2}. Let us calculate the  bounadry value conditions for an $n^{\text{th}}$ NB barrier located at $x_n=0$. This boundary condition is between $n^{\text{th}}$ N region and $n^{\text{th}}$ B region. The wavefunction for this is given in \eqaref{nN} and \eqref{nB}.
\begin{widetext}
    \begin{subequations} \label{bvcbarrier1}
    \begin{align} 
    a_n \times e^{iK_1 (n-1)(D+d) }\chi_{\varepsilon_+}(-X) + b_n \times 0 &= a_{B,n} \times e^{iK_1 ((n-1)(D+d)+d) } \chi_{\varepsilon_{B+}}(-X) + b_{B,n} \times 0 \\
    a_n \times 0 + b_n \times e^{iK_2 (n-1)(D+d) } \chi_{\varepsilon_-}(-X) &= a_{B,n} \times 0 + b_{B,n} \times e^{iK_2 ((n-1)(D+d)+d) }\chi_{\varepsilon_{B-}}(-X) \\
    a_n \times e^{iK_1 (n-1)(D+d) } \at{\derivative{\chi_{\varepsilon_+}}{x}}{x=0} + b_n \times 0 &= a_{B,n} e^{iK_1 ((n-1)(D+d)+d) } \times \at{\derivative{\chi_{B,\varepsilon_+}}{x}}{x=0} + b_{B,n} \times 0 \\
    a_n \times 0 + b_n \times e^{iK_2 (n-1)(D+d) } \at{\derivative{\chi_{\varepsilon_-}}{x}}{x=0} &= a_{B,n} \times 0 + b_{B,n} \times \times e^{iK_2 ((n-1)(D+d)+d) } \at{\derivative{\chi_{B,\varepsilon_-}}{x}}{x=0}
\end{align}
\end{subequations}
\end{widetext}
From this we calculate the matrix $M_{B,lat}^{(n)}$ defined in \eqaref{barrierM} in main text. 

\section{Lifting of degeneracy in dispersion with single barrier in the quantum Hall region} \label{degap}

Now let us discuss the single barrier dispersion shown in \figref{disp1} (a). For a square barrier inside the quantum Hall region, inside the barrier $H_0$ from \eqaref{hamfg} becomes,

\begin{equation} \label{hamb}
    H_0(x,X) =-\frac{1}{2 } \frac{d^2}{dx^2} + \frac{1}{2} (x - X)^2 +V_0 \Theta (-x+\frac{d}{2}) \Theta(x+\frac{d}{2})
\end{equation}

Now, the axis transformation to create the $M_{NB}$ (given in \eqaref{bigM} in the main text) matrix, $X\rightarrow X+ d/2$ (discussed in the main text). The boundary value conditions are discussed in \eqaref{bvcbarrier} in \apref{ap2}. Now \eqaref{hamb} becomes,

\begin{equation} \label{hamf}
    H_0(x,X) =-\frac{1}{2 } \frac{d^2}{dx^2} + \frac{1}{2} (x - X -\frac{d}{2})^2 +V_0 \Theta (-x+\frac{d}{2}) \Theta(x+\frac{d}{2}).
\end{equation}

As we are discussing the matrix $M_{NB}$ due to boundary condition at $x=-d/2$, (left side of the barrier), the value of the second term in the Hamiltonian becomes $\frac{1}{2}(X+d)^2$ at $x=-d/2$. We solve the eigenvalue equation for \eqref{hamb} and then match the boundary value conditions at $x=-d/2$ to obtain $M_{NB}$. The dispersion relation is plotted as $E$ vs $X$. The effect of the term $\frac{1}{2}(X+d)^2$ can be seen at $X=d$ in the dispersion plot. For a given value of energy, $E$, LLs of different orders form in the neighbourhood of the boundary and correspond to different values of $X$ near the boundary. In our case the NB boundary is at $x=-d$. Hence the lifting of degeneracy due to the barrier potential are shifted in the $X$ axis as compared to exact edges of NB boundary. Comparing with the case of a confined quantum Hall system \cite{Halperin1982} we also see the exact location depends on the quantum number of that LL.
In \figref{SNS} (b) we show the location where the lifting of degeneracy occurs in the LLs. In the case considered here, $d=2$. Hence we see that this location for $f_X$ LLs occur in the neighbourhood of the point $X=2$ in the dispersion plot.


Now let us discuss the lifting of degeneracy inside the barrier region with the help of \figref{SNS} (b). The eigenvalues inside the barrier from \eqaref{hamf} can be written as $E_{b,e}= \left[ (n_e+ 1/2)  - \nu/2 + V_0 \right]$. For hole states $g_X$, the eigenvalues are,
$E_{b,h}= -\left[ (n_e+ 1/2)  - \nu/2 + V_0 \right]$. This shows that the change in eigenvalue is opposite for $f_X$ LLs and $g_X$ LLs. Due to the presence of the barrier, we now have two regions of degenerate states, (i) inside the barrier and (ii) outside the barrier. The boundary condition allows the energy to continuously change from $\left[ (n_e+ 1/2)  - \nu/2 \right]$ to  $ \left[ (n_e+ 1/2)  - \nu/2 + V_0 \right]$ and form intermediate chiral edge states (ICES), whose role in transport forms the bedrock of this work. The same analysis can be done for the $g_X$ LLs and their direction is opposite to the $f_X$ LLs in the dispersion plot.

\section{Distribution of various current components}\label{AppI}
The total quasiparticle charge current in the superconductor-quantum Hall juctions is given by \cite{Hoppe2000}

\begin{equation} \label{chcr}
	I_{X}^{(Q)} = I_{X}^{(Q,n)} - I_{X}^{(Q,a)} + I_{X}^{(Q,s)} .
\end{equation}

This current captures the contribution from the currect carrying edge states (normal reflection) and Andreev reflection in such junctions. It has three components which composes ordinary edge current $I_{X}^{(Q,n)}$, Andreev reflection contribution ($I_{X}^{(Q,a)}$) and Supercurrent ($I_{X}^{(Q,s)}$).

\begin{subequations} \label{chcurr}
	\begin{align}
		I_{X}^{(Q,n)} & = \frac{e l^2}{\hbar L_y} \frac{\partial E }{\partial X} \\
		I_{X}^{(Q,a)} & = \frac{e l^2}{\hbar L_y} \frac{\partial E }{\partial X} 2 \int_{x} \abs{g_x(x)}^2 \\
		I_{X}^{(Q,s)} & = \frac{e l^2}{\hbar L_y} 2 \Delta \int_{x} \Theta(-x) \left[g_X^* \frac{df_X}{dX} - f_X^* \frac{dg_X}{dX} \right]
	\end{align}
\end{subequations}

\section{Graphene Based S-QH-S junction}\label{apsgs}
The Dirac-BdG (DBdG) equation for a monolayer-graphene based SQHS junction in the presence of an uniform magnetic field is given by (details in \cite{Parthathesis}),
\begin{widetext}
    \begin{equation}
	\begin{bmatrix}
		v_F \tau_0 \otimes (\mathbf{p}+ e \mathbf{A})\vdot \mathbf{\sigma} + V(x) - \mu && \Delta \\
		\Delta^* && \mu - v_F \tau_0 \otimes (\mathbf{p}- e \mathbf{A})\vdot \mathbf{\sigma} -V(x)
	\end{bmatrix} \begin{bmatrix}
		\Psi_e\\
		\Psi_h
	\end{bmatrix}= \varepsilon \begin{bmatrix}
		\Psi_e\\
		\Psi_h
	\end{bmatrix} \label{GS}
\end{equation}
\end{widetext}

\subsection{Solution in the G Region}

In the absence of any external potential, G region the DBdG equation becomes,

\begin{widetext}
    \begin{equation}
	\begin{bmatrix}
		v_F \tau_0 \otimes (\mathbf{p}+ e \mathbf{A})\vdot \mathbf{\sigma}  - \mu && 0 \\
		0 && \mu - v_F \tau_0 \otimes (\mathbf{p}- e \mathbf{A})\vdot \mathbf{\sigma}
	\end{bmatrix} \begin{bmatrix}
		\Psi_e\\
		\Psi_h
	\end{bmatrix}= \varepsilon \begin{bmatrix}
		\Psi_e\\
		\Psi_h
	\end{bmatrix} \label{G1}
\end{equation}
\end{widetext}

Here the wavefunctions contain $\Psi_e$ and $\Psi_h$, a pair of four dimensional vectors, which represent the electron and hole excitaion. If we choose $\Psi_e$ and $\Psi_h$ as $\Psi_e=(\psi_{e1} , \psi_{e2}, \psi_{e3}, \psi_{e4})$ and $\Psi_h=(\psi_{h1},\psi_{h2},\psi_{h3},\psi_{h4})$, then from \eqaref{G1}, we can write

\begin{widetext}
    \begin{equation} \label{sgseh}
	\begin{bmatrix}
		 - \mu && v_F (\pi_x- i \pi_y) && 0 && 0 \\
		v_F (\pi_x+ i \pi_y) && - \mu  && 0 &&  0 \\
	0 && 0 && \mu  && - v_F (\bar{\pi}_x-i \bar{\pi}_y)\\
		0 && 0 && - v_F (\bar{\pi}_x+i \bar{\pi}_y) && \mu 
	\end{bmatrix} \begin{bmatrix}
		\psi_{e1}\\ \psi_{e2}\\ \psi_{h1}\\ \psi_{h2}
	\end{bmatrix}= \varepsilon \begin{bmatrix}
		\psi_{e1}\\ \psi_{e2}\\ \psi_{h1}\\ \psi_{h2} 
	\end{bmatrix}
\end{equation}
\end{widetext}

and,

\begin{widetext}
    \begin{equation}
	\begin{bmatrix}
		 - \mu && v_F (\pi_x- i \pi_y) && 0 && 0 \\
		v_F (\pi_x+ i \pi_y) && - \mu  && 0 &&  0 \\
		0 && 0 && \mu  && - v_F (\bar{\pi}_x-i \bar{\pi}_y)\\
		0 && 0 && - v_F (\bar{\pi}_x+i \bar{\pi}_y) && \mu 
	\end{bmatrix} \begin{bmatrix}
		\psi_{e3}\\ \psi_{e4}\\ \psi_{h3}\\ \psi_{h4}\\
	\end{bmatrix}= \varepsilon \begin{bmatrix}
		\psi_{e3}\\ \psi_{e4}\\ \psi_{h3}\\ \psi_{h4}\\
	\end{bmatrix} \label{32}
\end{equation}
\end{widetext}

Let us first start with \eqaref{sgseh}, the electron and hole parts of this equation can be decoupled. From the electron part,

\begin{subequations} \label{G2}
	\begin{align}
		- \mu	\psi_{e1}+v_F (\pi_x- i \pi_y)\psi_{e2} &= \varepsilon \psi_{e1} \\
		v_F (\pi_x+ i \pi_y)  \psi_{e1} - \mu \psi_{e2} &= \varepsilon \psi_{e2}
	\end{align}
\end{subequations}

In case of uniform magnetic field, we have,

\begin{equation}
	\mathbf{A}= \begin{cases}
		& B x \hat{y} \text{ , G region} \\
		& 0 \text{ , S region}
	\end{cases}
\end{equation}

If we take $k_y=\frac{p_{y}}{\hbar}=\frac{X}{l^2}$, then from \eqaref{G2}, we get,

\begin{equation} \label{G3}
	\left[\hat{p_x}^2 + \hbar eB + \left(\hbar \frac{X}{l^2} + eBx\right)^2\right] \psi_{e1}= \left(\frac{\varepsilon+ \mu}{v_F}\right)^2 \psi_{e1}
\end{equation}

Now, $\left(\hbar \frac{X}{l^2} + eBx\right) = \sqrt{\hbar e B} \left[\frac{X}{l} + \frac{x}{l}\right]$, where, $l= \sqrt{\frac{\hbar}{eB}}$ is the magnetic length. If we now measure length in the units of $l$ and energy in the units of $\hbar v_F/l$,

\begin{align*}
	x & \rightarrow x l \\
	X & \rightarrow X l \\
	(\varepsilon+\mu) & \rightarrow (\varepsilon+\mu) \times \frac{\hbar v_F}{l} 
\end{align*}

This makes \eqaref{G3} and the corresponding equation of $\psi_{h1}$ describing the hole part,

\begin{subequations} \label{Greg}
    \begin{equation} \label{psie1}
	\left[-\frac{1}{2} \partial_x^2 + \frac{1}{2}(x+X)^2\right]\psi_{e1} = \frac{1}{2}\left[( \mu + \varepsilon)^2-1\right]\psi_{e1}
\end{equation}
\begin{equation}
    \left[-\frac{1}{2} \partial_x^2 + \frac{1}{2}(x-X)^2\right]\psi_{h1} = \frac{1}{2}\left[( \mu -\varepsilon  )^2-1\right]\psi_{h1}
\end{equation}
\end{subequations}


The solutions of $\psi_{e1}$ from \eqaref{psie1}  of this equation is

\begin{align}
	\psi_{e1} &=-i \left(\varepsilon+ \mu \right) e^{- \frac{1}{2} (x+X)^2} H_{\frac{1}{2} \left(\varepsilon+ \mu \right)^2 -1 } (x+X) \\
	\psi_{e2} &= e^{- \frac{1}{2} (x+X)^2} H_{\frac{1}{2} \left(\varepsilon+ \mu \right)^2 } (x+X)
\end{align}

The wavefunctions in the G region is written as,

\begin{equation} \label{wfg}
	\bm{\Psi}(x,y)= e^{ik_y y} \mqty(C_e\otimes\Phi_e(x+X) \\
							 C_h\otimes\Phi_h(x-X) )
\end{equation}

where, \begin{equation}
	\Phi_e (\xi)= e^{-(1/2)\xi^2} \mqty( -i(\mu+\varepsilon) H_{\frac{1}{2} \left(\varepsilon+ \mu \right)^2 -1 }(\xi) \\
	H_{\frac{1}{2} \left(\varepsilon+ \mu \right)^2 } (\xi)	)
\end{equation}

\begin{equation}
	\Phi_h (\xi)= e^{-(1/2)\xi^2} \mqty(  H_{\frac{1}{2} \left(\mu -\varepsilon \right)^2}(\xi) \\
	-i(\mu- \varepsilon) H_{\frac{1}{2} \left(\mu -\varepsilon \right)^2 -1 } (\xi)	)
\end{equation}

If we restrict our calculation to single valley, then the wavefunction $ \Psi(x,y) =  \begin{bmatrix}
	\psi_{e1} & \psi_{e2} & \psi_{h1} & \psi_{h2}
\end{bmatrix}^T$ has two solutions in the graphene region.
\begin{subequations} \label{grwf}
    \begin{equation}
	\Psi_{G1} =  e^{ik_y y} e^{-(1/2)(x+X)^2} \mqty( -i(\mu+\varepsilon) H_{\frac{1}{2} \left(\varepsilon+ \mu \right)^2 -1 }(x+X) \\
	H_{\frac{1}{2} \left(\varepsilon+ \mu \right)^2 } (x+X) \\
	0 \\
	0	)
\end{equation}

\begin{equation}
	\Psi_{G2} =  e^{ik_y y} e^{-(1/2)(x-X)^2} \mqty( 
	0 \\
	0\\
	H_{\frac{1}{2} \left(\mu -\varepsilon \right)^2}(x-X) \\
	-i(\mu- \varepsilon) H_{\frac{1}{2} \left(\mu -\varepsilon \right)^2 -1 } (x-X)	
	)
\end{equation}
\end{subequations}

Here $H_{\alpha}(\xi)$ is Hermite polynomial of degree $\alpha$.

	\subsection{Solution in the Barrier region}
	
	In the earlier calculation of S-QH-S junctions in non relativistic two dimensional electronic system, we had taken the barrier potential in the units of $\hbar \omega_C$. Here we take the energies in the units of $\frac{\hbar v_F}{l}$. We have also scaled the external scattering potential accordingly. If we put a electrostatic potential barrier of height $V_0$  in the graphene region, then we can write the DBdG equation as 
	
	\begin{widetext}
	    \begin{equation}
		\begin{bmatrix}
			V_0 - \mu && v_F (\pi_x- i \pi_y) && 0 && 0 \\
			v_F (\pi_x+ i \pi_y) && V_0 - \mu  && 0 &&  0 \\
			0 && 0 && \mu -V_0  && - v_F (\bar{\pi}_x-i \bar{\pi}_y)\\
			0 && 0 && - v_F (\bar{\pi}_x+i \bar{\pi}_y) && \mu -V_0
		\end{bmatrix} \begin{bmatrix}
			\psi_{e1}\\ \psi_{e2}\\ \psi_{h1}\\ \psi_{h2}
		\end{bmatrix}= \varepsilon \begin{bmatrix}
			\psi_{e1}\\ \psi_{e2}\\ \psi_{h1}\\ \psi_{h2} \label{31a}
		\end{bmatrix}
	\end{equation}
	\end{widetext}

The counterparts of equations \eqref{Greg} becomes,
\begin{subequations} 
    \begin{equation} 
	\left[-\frac{1}{2} \partial_x^2 + \frac{1}{2}(x+X)^2\right]\psi_{e1} = \frac{1}{2}\left[( \mu + \varepsilon -V_0)^2-1\right]\psi_{e1}
\end{equation}
\begin{equation}
    \left[-\frac{1}{2} \partial_x^2 + \frac{1}{2}(x-X)^2\right]\psi_{h1} = \frac{1}{2}\left[( \mu  -\varepsilon - V_0 )^2-1\right]\psi_{h1}
\end{equation}
\end{subequations}

With this, the modified solutions in this region become,

\begin{widetext}
    \begin{subequations} \label{bwf}
    \begin{equation}
	\Psi_{B1}(x,y) =  e^{ik_y y} e^{-(1/2)(x+X)^2} \mqty( -i(\mu+\varepsilon - V_0) H_{\frac{1}{2} \left(\varepsilon - V_0 + \mu \right)^2 -1 }(x+X) \\
	H_{\frac{1}{2} \left(\varepsilon - V_0 + \mu \right)^2 } (x+X) \\
	0 \\
	0	)
\end{equation}

\begin{equation}
	\Psi_{B2} (x,y) =  e^{ik_y y} e^{-(1/2)(x-q)^2} \mqty( 
	0 \\
	0\\
	H_{\frac{1}{2} \left(\mu - V_0 -\varepsilon \right)^2}(x-q) \\
	-i(\mu - V_0- \varepsilon) H_{\frac{1}{2} \left(\mu - V_0 -\varepsilon \right)^2 -1 } (x-q)	
	)
\end{equation}
\end{subequations}
\end{widetext}

	\subsection{Solution in Superconducting Region}

In the superconducting region, we can write

	\begin{widetext}
	\begin{equation}
		\begin{bmatrix}
			- \mu -U_0 && v_F (p_x- i p_y) &&  \Delta_0 e^{i \Phi} && 0 \\
			v_F (p_x+ i p_y) && - \mu -U_0 && 0 &&  \Delta_0 e^{i \Phi} \\
			\Delta_0 e^{-i \Phi} && 0 && \mu +U_0  && - v_F (p_x-i p_y)\\
			0 && \Delta_0 e^{-i \Phi} && - v_F (p_x+i p_y) && \mu  +U_0
		\end{bmatrix} \begin{bmatrix}
			\psi_{e1}\\ \psi_{e2}\\ \psi_{h1}\\ \psi_{h2}
		\end{bmatrix}= \varepsilon \begin{bmatrix}
			\psi_{e1}\\ \psi_{e2}\\ \psi_{h1}\\ \psi_{h2} 
		\end{bmatrix} \label{supl}
	\end{equation}
\end{widetext}

Here, $\Phi$ is the superconducting phase. The possible solutions in the regime of $U_0+\mu \gg \Delta_0, \varepsilon$ are given by

\begin{subequations} \label{super}
	\begin{equation}
		\psi_{S1} (\Phi)=e^{ik_y y+ik_ox-k_i x}\left(
		\begin{array}{ccc}
			e^{i\beta} \\ 	
			e^{i\beta+i\gamma} \\ 
			e^{-i\Phi} \\
			e^{-i\Phi+i\gamma} \\
		\end{array} 
		\right),
	\end{equation}
	\begin{equation}
		\psi_{S2} (\Phi)=e^{ik_y y+ik_ox+k_i x}\left(
		\begin{array}{ccc}
			e^{-i\beta} \\ 	
			e^{-i\beta+i\gamma} \\ 
			e^{-i\Phi} \\
			e^{-i\Phi+i\gamma} \\
		\end{array} 
		\right),
	\end{equation}
	\begin{equation}
		\psi_{S3} (\Phi)=e^{ik_y y-ik_ox-k_i x}\left(
		\begin{array}{ccc}
			e^{-i\beta} \\ 	
			-e^{-i\beta-i\gamma} \\ 
			e^{-i\Phi} \\
			-e^{-i\Phi-i\gamma} \\
		\end{array} 
		\right),
	\end{equation}
	\begin{equation}
		\psi_{S4} (\Phi)=e^{ik_y y-ik_ox+k_i x}\left(
		\begin{array}{ccc}
			e^{i\beta} \\ 	
			-e^{i\beta-i\gamma} \\ 
			e^{-i\Phi} \\
			-e^{-i\Phi-i\gamma} \\
		\end{array} 
		\right).
	\end{equation}
\end{subequations}

where, 

\begin{equation} \label{16}
	\beta =
	\left\{
	\begin{array}{ll}
		\cos^{-1}\left( \frac{\varepsilon}{\Delta_o} \right)  & \mbox{if } \varepsilon < \Delta_o \\
		-i \cosh^{-1}\left( \frac{\varepsilon}{\Delta_o} \right) & \mbox{if } \varepsilon > \Delta_o
	\end{array}
	\right.
\end{equation}

\begin{equation} \label{51}
	\gamma= \sin^{-1}\left[ \frac{\hbar v_F k_y }{U_o+\mu} \right]
\end{equation}

\begin{equation} \label{52}
	k_o=\sqrt{\left(\frac{U_o+\mu}{\hbar v_F}\right)^2 -k_y^2}
\end{equation}
\begin{equation} \label{53}
	k_i=\frac{(U_o+\mu)\Delta_o}{\hbar^2 v_F^2 k_o}\sin \beta
\end{equation}

In the regime of $|k_y|\le \frac{\mu}{\hbar v_F}$ and if we take $U_o\gg \mu ,\varepsilon$ then, $\gamma \rightarrow 0$, $k_o \rightarrow \frac{U_o}{\hbar v_F}$ and $k_i \rightarrow \frac{\Delta_o}{\hbar v_F} \sin \beta $. In the region $x<0$, (\ie left superconductor), the wavefunction is $\Psi_l= a_1 \psi_2 (\phi_1) + a_2 \psi_4 (\phi_1) $. In the region $x>L$, (\ie right superconductor) the wave-function is $\Psi_{r}= b_1 \psi_1(\phi_2) +	b_2 \psi_3(\phi_2)$.

\subsection{Boundary Value Condition} \label{bvcgraphene}

The transfer matrices are calculated using  boundary value conditions in the same method we used earlier in set of equations \eqaref{bvcq1} and \eqref{bvcbarrier} in \apref{ap2}.
We match the wavefunctions \eqaref{super} and \eqref{grwf}at the SG interface to get $M_{SG}$, the wavefunctions \eqaref{grwf} and \eqref{bwf} at the GB interface to get $M_{GB}$,  the wavefunctions \eqaref{bwf} and \eqref{grwf} at the BG interface to get $M_{BG}$ and  the wavefunctions \eqaref{grwf} and \eqref{super} at the GS interface to get $M_{GS}$. Now we construct the transfer matrix similar to \eqaref{bigM}

\begin{equation} \label{bigMG}
	\mathcal{M}= \begin{bmatrix}
		M_{SG} && 0 && 0 && 0 \\
		0 && M_{GB} && 0 && 0 \\
		0 && 0 && M_{BG} && 0 \\
		0 && 0 && 0 &&  M_{GS}
	\end{bmatrix}.
\end{equation}

We are not providing the explicit form of these matrices. As the graphene wavefunctions have four component, all the matrices in \eqaref{bigMG} are $(4 \times 4)$ matrices. The $E$ - $X$ dispersion in this case is given by
\beq \label{MG}
det(\mathcal{M}) = det(M_{SG}) det(M_{GB})det(M_{BG})det(M_{GS}) = 0
\eeq 

\section{Table of Hole Probabilities} \label{tabhole}
Please note that the significant digit in the values given in the table below, is purely a computational artifact.
\begin{table}[h!]
	\centering
	
	\begin{minipage}{0.45\textwidth}
		\centering
		\begin{tabular}{|c|c|}
			\hline
			\textbf{n} & \textbf{$B_n$} \\ \hline \hline
			1 & 0.471016 \\ \hline
			2 & 0.499073 \\ \hline
			3 & 0.502286 \\ \hline
			4 & 0.365599 \\ \hline
			5 & 0.497525 \\ \hline
			6 & 0.500802 \\ \hline
			7 & 0.070487 \\ \hline
			8 & 0.528958 \\ \hline
		\end{tabular}
		\caption{Hole probabilities of $\nu=5.31$}
        \label{tab1}
	\end{minipage}
	\hfill
	\begin{minipage}{0.45\textwidth}
		\centering
		\begin{tabular}{|c|c|}
			\hline
			\textbf{n} & \textbf{$B_n$} \\ \hline \hline
			1 & 0.465506 \\ \hline
			2 & 0.500595 \\ \hline
			3 & 0.501658 \\ \hline
			4 & 0.498537 \\ \hline
			5 & 0.499180 \\ \hline
			6 & 0.534521 \\ \hline
		\end{tabular}
		\caption{Hole probabilities of $\nu=5.41$}
        \label{tab2}
	\end{minipage}
\end{table}

\begin{table}[h!]
	\centering
	\begin{tabular}{|c|c|c|c|}
		\hline
		n & $B_n$ for $\nu/2 = 3.555$ & $B_n$ for $\nu/2 = 3.560$ & $B_n$ for $\nu/2 = 3.565$\\ \hline \hline
		1 & 0.50275 & 0.48505 & 0.47407\\ \hline
		2 & 0.49439 & 0.49630 & 0.49795\\ \hline
		3 & 0.50486 & 0.50405 & 0.50315\\ \hline
		4 & 0.49543 & 0.49607 & 0.49661\\ \hline
		5 & 0.50456 & 0.50418 & 0.50339\\ \hline
		6 & 0.00230 & 0.00004 & 0.07291\\ \hline
		7 & 0.49509 & 0.49586 & 0.49669\\ \hline
		8 & 0.50566 & 0.58670 & 0.50193\\ \hline
		9 & 0.48991 & 0.50387 & 0.32144\\ \hline
		10 & 0.49729 & 0.51516 & 0.52569\\ \hline
	\end{tabular}
    \caption{Hole probabilities of $\nu=3.555$, $3.560$ and $3.565$}
    \label{tab3}
\end{table}

		\bibliography{reference}
	\end{document}